\documentclass[a4paper,11pt]{article}
\usepackage[top=3cm,bottom=3.4cm,left=2cm,right=2cm]{geometry}

\usepackage{color}
\usepackage{bm}
\usepackage{amsmath}
\usepackage{amssymb}
\usepackage{cite}
\usepackage[dvipdfmx]{graphicx}
\DeclareFontFamily{OT1}{pzc}{}
\DeclareFontShape{OT1}{pzc}{m}{it}{<-> s * [1.10] pzcmi7t}{}
\DeclareMathAlphabet{\mathpzc}{OT1}{pzc}{m}{it}
\newcommand{\nn}{\nonumber\\}
\newcommand{\bra}[1]{\!\left<#1 \right|}
\newcommand{\ket}[1]{\left| #1 \right>\!}

\renewcommand{\thepage}{}
\makeatletter
\@addtoreset{equation}{section}
\renewcommand{\theequation}{\thesection.\@arabic\c@equation}
\makeatother
\renewcommand{\thefootnote}{\fnsymbol{footnote}}

\begin{document}
\begin{titlepage}
\title{
\vspace*{-4ex}
\hfill
\begin{minipage}{3.5cm}
\end{minipage}\\
 \bf 
Closed string vertex operators \\
with various ghost number
\vspace{0.5em}
}

\author{
Isao~{\sc Kishimoto},$^{1}$\footnote{\tt ikishimo@rs.socu.ac.jp}
~~~Mako {\sc Kouga},$^{2}$\footnote{\tt kouga@asuka.phys.nara-wu.ac.jp}
~~~Shigenori {\sc Seki}$^{3,4}$\footnote{\tt sigenori@yukawa.kyoto-u.ac.jp}
\\
\vspace{-3ex}\\
~and~ 
\\
\vspace{-3ex}\\
Tomohiko~{\sc Takahashi}$^{2}$\footnote{\tt tomo@asuka.phys.nara-wu.ac.jp}
\\
\vspace{0ex}\\
\\
$^{1}${\it
 Center for Liberal Arts and Sciences, Sanyo-Onoda City University,}\\
{\it Daigakudori 1-1-1, Sanyo-Onoda Yamaguchi 756-0884, Japan}
\vspace{1ex}
\\
$^{2}${\it Department of Physics, Nara Women's University,}\\
{\it Nara 630-8506, Japan}
\vspace{1ex}
\\
$^{3}${\it Osaka Central Advanced Mathematical Institute (OCAMI),}\\
{\it Osaka Metropolitan  University,}\\
{\it 3-3-138, Sugimoto,
 Sumiyoshi-ku, Osaka 558-8585, Japan}
\vspace{1ex}
\\
$^{4}${\it Institute for Fundamental Sciences, Faculty of Science and Engineering,}\\
{\it  Setsunan University,}\\
{\it  17-8, Ikedanaka-machi, Neyagawa, Osaka 572-8508, Japan}\\
\vspace{0ex}
}

\date{}
\maketitle

\begin{abstract}
\normalsize

We construct closed string vertex operators with various ghost numbers
in addition to the conventional ones, using the Faddeev-Popov procedure
for the gauge fixing of the conformal Killing group, from matter primary
fields. We find that these operators give solutions to the descent
equations in the framework of the BRST formalism. Similarly, we also
construct solutions to the descent equations for the dilaton vertex
operator with the Lorentz covariant form. Using the unintegrated vertex
operator of the dilaton with the ghost number three, we obtain the
correct result of the tadpole amplitude on the disk, including a
non-zero contribution from a BRST exact term which comes from a
conformal transformation.

\end{abstract}

\end{titlepage}

\renewcommand{\thepage}{\arabic{page}}
\renewcommand{\thefootnote}{\arabic{footnote}}
\setcounter{page}{1}
\setcounter{footnote}{0}


\section{Introduction}

In the BRST formalism, physical states are defined as those that are
annihilated by the BRST charge, and the physical observables are
obtained by computing the cohomology of the BRST
charge\cite{Kugo:1979gm}.  In bosonic closed string theory, the BRST
charge $Q$ is given by the sum of the left and right-moving BRST
charges, $Q_{\rm B}$ and $\tilde{Q}_{\rm B}$, and the physical states
$\left|\mathrm{phys}\right>$ are then defined as the states which satisfy
\cite{Kato:1982im}
\begin{eqnarray}
 Q\left|\mathrm{phys}\right>=0.
\label{eq:KO}
\end{eqnarray}
In addition to the BRST charge, there is a constraint known as 
the level-matching condition imposed on the closed string states:
\begin{eqnarray}
 (L_0-\tilde{L}_0)\left|\mathrm{phys}\right>=0,
\label{eq:L0phys}
\end{eqnarray}
where $L_n$ and $\tilde{L}_n$ are the left and right-moving components
of the total Virasoro generators, respectively.
In the BRST formalism, this is usually replaced by imposing
the equivalent condition
\begin{eqnarray} 
(b_0-\tilde{b}_0)\left|\mathrm{phys}\right>=0,
\label{eq:b0phys}
\end{eqnarray}
where $b_n$ and $\tilde{b}_n$ are the left and right-moving modes of
anti-ghost fields, $b(z)$ and $\tilde{b}(\bar{z})$.  Two conditions
(\ref{eq:KO}) and (\ref{eq:b0phys}) provide semi-relative BRST
cohomology, which is extensively used to investigate string amplitudes
\cite{Nelson:1988ic,La:1989xk,Distler:1990ea,Becchi:1993pp}.
Moreover, conventionally, 
covariant closed string field theory is formulated by
the string field constrained by the level-matching condition.

However, since the early days of closed string theory, the absolute BRST
cohomology has been given by
\begin{eqnarray}
 \left|\mathrm{phys}\right>=
\left|\mathrm{P}_1\right>\otimes \tilde{c}_1 c_1\ket{0}
+\left|\mathrm{P}_2\right>\otimes c_0\tilde{c}_1 c_1\ket{0}
+\left|\mathrm{P}_3\right>\otimes \tilde{c}_0\tilde{c}_1 c_1\ket{0}
+\left|\mathrm{P}_4\right>\otimes \tilde{c}_0c_0\tilde{c}_1 c_1\ket{0}
+Q\left|\chi\right>,
\label{eq:abscohomology}
\end{eqnarray}
where $c_n$ and $\tilde{c}_n$ are the left and right-moving modes of
ghost fields, $c(z)$ and $\tilde{c}(\bar{z})$, and $\left|\mathrm{P}_i\right>$ are
transverse physical states in the matter sector \cite{Henneaux:1986kp}.
It should be noticed that the absolute cohomology includes states which
do not satisfy the constraint (\ref{eq:b0phys}).  In addition, despite
only the condition (\ref{eq:KO}) being imposed, the cohomology
automatically satisfies the level-matching condition (\ref{eq:L0phys}).
This result implies that the level-matching condition does not
necessarily have to be imposed on physical states in the BRST formalism.
Actually, some recent works investigate to construct closed string field
theories without the level-matching condition
\cite{Erbin:2022cyb,Okawa:2022mos}.

The absolute cohomology (\ref{eq:abscohomology}) includes states with
ghost zero modes, and transverse states appear quadruply within the
cohomology. Among the four sectors, cohomology proportional to $c_0^+\
(=(c_0+\tilde{c}_0)/2)$ is believed to be a BRST exact state.  The state
$c_0^+\ket{\psi}$, satisfying $Qc_0^+\ket{\psi}=0$, can be formally
expressed as $c_0^+\ket{\psi}=Q((b_0+\tilde{b}_0)/(L_0+\tilde{L}_0)\,
c_0^+\ket{\psi})$.  In the case being on-shell, this state becomes
ill-defined due to $L_0+\tilde{L}_0$ being zero on
$\ket{\psi}$. However, as discussed in \cite{Thorn1989},\footnote{See
section 2.3 in \cite{Thorn1989}.} using the continuity of
$L_0+\tilde{L}_0$, the state can be expressed as BRST exact through a
limit approached from the off-shell region.  Similarly, the avoidance of
multiplicity is discussed in \cite{Henneaux:1986kp},
using the relativistic spinless particle as an example. Indeed,
with these methods, it is easy to derive
a BRST exact expression for certain on-shell states.

What is crucial here, as stated in \cite{Thorn1989}, is that the argument
using the continuity above cannot be applied to the cohomology
proportional $c_0^-\ (=(c_0-\tilde{c}_0)/2)$.
This is due to the discrete spectrum of $L_0-\tilde{L}_0$,
which arises from constraints on the left and right momenta.
Accordingly, only from the perspective of BRST cohomology, we have no reason to
exclude states that include $c_0^-$---which does not conform to 
(\ref{eq:b0phys})---from the sets of physical states.

In this paper, we explore the possibility that the level-matching
condition may not be as fundamental as previously thought in the BRST
formalism and investigate closed string vertex operators with various
ghost numbers, which are suggested to exist from the absolute BRST
cohomology (\ref{eq:abscohomology}) and the state-operator
correspondence. 

The vertex operator for a closed string state is given
by
\begin{eqnarray}
 \int d^2 z V(z,\bar{z}),
\label{eq:gh0op}
\end{eqnarray}
where $V(z,\bar{z})$ is a matter primary operator of weight $(1,1)$. The
scattering amplitude is given by the Polyakov path integral with the
insertions of these vertex operators integrated over a world-sheet.
Conventionally, the positions of some
vertex operators are fixed to eliminate the gauge redundancy 
and the operators are multiplied
by ghost fields:\footnote{See for example the book \cite{Polchinski:1998rq}
by J.~Polchinski.}
\begin{eqnarray}
 \tilde{c}(\bar{z})c(z)V(z,\bar{z}).
\label{eq:gh2op}
\end{eqnarray}
For fixed positions, $dz\ (\,d\bar{z}\,)$ in the integrated vertex operator
is formally replaced by $c(z)\ (\,\tilde{c}(\bar{z})\,)$
and then the operator increases the ghost number by two.
Namely, differentials are related to ghost fields:
\begin{eqnarray}
&&dz~~\leftrightarrow~~ c(z),
\qquad
d\bar{z}~~\leftrightarrow~~ \tilde{c}(\bar{z}).
\label{eq:dzc}
\end{eqnarray}
It is noted that the operator (\ref{eq:gh2op}) corresponds to the state of the
first term in the absolute BRST cohomology (\ref{eq:abscohomology}).

Following the correspondence of (\ref{eq:dzc}), it is possible to
construct the closed string vertex operator with ghost number one. For
instance, in the case that we fix the real part of $z=x+iy$ and leave
the integration along the imaginary axis, $dx=(dz+d\bar{z})/2$
is replaced by
$(c+\tilde{c})/2$ and then the vertex operator is given by\footnote{We
use the definition $d^2z=2dxdy$.}
\begin{eqnarray}
 \int dy\,(c(z)+\tilde{c}(\bar{z}))V(z,\bar{z}).
\label{eq:gh1op}
\end{eqnarray}
This operator is given also by acting an anti-ghost insertion associated
with the $y$ integration on the operator (\ref{eq:gh2op}). The latter
is a conventional procedure for a moduli integration and so this
operator itself is not novel.

Even if (\ref{eq:dzc}) is useful, it is impossible to reach the vertex
operator with ghost number three, which corresponds to the state of the
second and third terms in (\ref{eq:abscohomology}). In this paper, we
will begin with a reconsideration of the gauge fixing of $PSL(2,\mathbb{R})$
symmetry in a disk amplitude with a closed string insertion, and as a
result, we will mainly propose that closed string vertex operators with various
ghost numbers can be constructed by multiplying the operator
\begin{eqnarray}
 \frac{i}{4\pi}(
 \partial c(z)-\bar{\partial}\tilde{c}(\bar{z}))
\label{eq:dghfac}
\end{eqnarray}
to the operators (\ref{eq:gh0op}), (\ref{eq:gh2op}), (\ref{eq:gh1op})
and so on. It is worth noting that the vertex operators with the factor 
(\ref{eq:dghfac}) correspond to states that include $c_0^-$.

Furthermore, we will explore these vertex operators with various ghost
numbers, employing the perspective of differential forms.
These vertex operators can be identified as operator-valued forms.
Subsequently, we will
establish that these forms can be related by descent equations with
respect to the BRST transformation. Additionally, we will demonstrate
that the supplementary terms incorporated into these forms during
conformal transformation can be expressed as a sum of total derivative
terms and BRST exact terms. Consequently, the amplitudes containing
these forms exhibit conformal invariance, even though the factor
(\ref{eq:dghfac}) may initially appear to contradict this property.

Finally, we will apply the advanced technique of the descent equation for
vertex operators to revisit the vertex operator associated with the
dilaton. It is worth noting that the simple correspondence between
(\ref{eq:gh0op}) and (\ref{eq:gh2op}) does not apply to the dilaton when
its momentum is zero. Consequently, we will successfully construct
vertex operators with various ghost numbers for the dilaton, including
the momentum-zero dilaton. In particular, the vertex operator of
dilaton with ghost number three represents a completely novel
expression, which facilitates the calculation of the dilaton tadpole in
the BRST formalism.

We will organize the paper as follows. In subsection \ref{sec:Inner}, we provide a
summary of our convention of vacuum, states, vertex
operators and disk or sphere amplitudes. Specifically, we adopt the
convention of treating the bra vacuum as the Hermitian conjugate of the
ket vacuum for a sphere correlator, even though the often-used
convention treats it as anti-Hermitian. In this subsection, we will
elucidate how this convention naturally establishes a correspondence
between an integrated and unintegrated vertex operator of closed
strings. In the rest of section \ref{sec:Closedstring}, we will examine closed string vertex
operators in the BRST formalism using the Faddeev-Popov (FP) procedure
to handle unconventional gauge fixing for the conformal Killing group (CKG) of
a disk, namely, $PSL(2,\mathbb{R})$. As a result, we will construct vertex
operators with different ghost numbers, and we will observe that these
operators include (\ref{eq:dghfac}) as a factor. 

In section \ref{sec:descent}, we treat
these vertex operators as operator-valued forms on a
world-sheet. We show that these forms satisfy the descent equations and
examine their conformal transformation, thus ensuring the conformal
invariance of amplitudes. From the descent equations, we will discover a
new vertex operator with ghost number one, which is not present in the
previous section.

In section \ref{sec:Dilaton}, we will build the vertex operators of
dilaton with various ghost numbers. An important distinction is that,
unlike other vertices, the dilaton vertex with ghost number three and
its ascendants incorporate the string coordinate $X^\mu$ directly,
rather than being derived from the derivative of $X^\mu$.  The inclusion
of $X^\mu$ appears to be necessary when dealing with vertices associated
with momentum zero. 

In section \ref{sec:Dilatontadpole}, we use the dilaton vertex with a ghost number of three to
compute the dilaton tadpole within the framework of the BRST
formalism. Just like with other vertices, the BRST exact term is
incorporated into the dilaton vertex during a conformal transformation.
This additional BRST exact term doesn't affect the amplitudes for most
vertex operators, but this isn't true for the dilaton vertex.  This is
because the correlation function involving BRST exact operators
constructed using $X^\mu$ may not necessarily equal zero, as discussed
in \cite{Belopolsky:1995vi}.  As a result, due to the non-zero
contribution of this additional term, it is essential to establish the
coordinate frame when defining the dilaton vertex initially.  From the
perspective of the state-operator correspondence, it is reasonable to
consider that the dilaton vertex resides at the origin of a unit disk.
Subsequently, by conformally mapping the unit disk to the upper-half
plane, we evaluate the dilaton tadpole as a correlation function in the
upper-half plane.  By incorporating the contribution of the additional
term, the resulting amplitude agrees with the established expression for
the tadpole in \cite{Liu:1987nz}.  It is worth noting that the dilaton
tadpole was originally derived in \cite{Douglas:1986eu}. Later, a subtle
point was addressed in \cite{Liu:1987nz}, emphasizing the necessity to
include the integration of the world-sheet curvature in the calculation
of the dilaton tadpole. In this paper, we compute the dilaton tadpole
consistently within the BRST formalism by dealing with the dilaton as an
unintegrated vertex operator.

Section \ref{sec:remarks} is devoted to concluding remarks.

\section{Closed string vertex operators with various ghost numbers
\label{sec:Closedstring}
}

\subsection{Inner products and unintegrated vertex operators
\label{sec:Inner}
}

Before proceeding further, let us examine the inner product of ground
states. To simplify our analysis, we will assume that the space-time is
flat Minkowski and all of the boundary conditions for the open string
are Neumann. We adopt the definition of $\left|0;k\right>$ in
\cite{Polchinski:1998rq} where the state is defined as the product of
the matter ground state and the ghost ground state, with momentum $k$
(see section 4.3 of \cite{Polchinski:1998rq}).  We can identify this
state with zero momentum as the following state in terms of the
conformal field theory vacuum and ghost oscillators:
\begin{eqnarray}
 \mathrm{open\ string:}&&~~~\left|0;0\right> = c_1\ket{0},
\\
  \mathrm{closed\ string:}&&~~~\left|0;0\right> = \tilde{c}_1c_1
\ket{0},
\end{eqnarray}
where $\ket{0}$ refers to the $SL(2,\mathbb{R})$ invariant vacuum state for open
strings or the $SL(2,\mathbb{C})$ invariant vacuum state for closed string.\footnote{
In the section 2.9 in \cite{Polchinski:1998rq}, the $SL(2,\mathbb{C})$ invariant state is represented as $\ket{1}$.}
Using the normalization of the ground states from
\cite{Polchinski:1998rq} (see (4.3.18a) and (4.3.18b) of
\cite{Polchinski:1998rq}), which respect hermiticity, the inner product
of these states can be expressed as\footnote{We adopt the following
hermiticity assignment:
\begin{eqnarray}
(c_m)^\dagger=c_{-m},~~~(\tilde{c}_m)^\dagger=\tilde{c}_{-m},~~~
(\,\ket{0}\,)^\dagger=\bra{0}.
\end{eqnarray}
}
\begin{eqnarray}
 \mathrm{open\ string:}&&~~~\bra{0}c_{-1} c_0 c_1 \ket{0}
=(2\pi)^{26}\delta^{26}(0),
\label{eq:openIP__0}
\\
  \mathrm{closed\ string:}&&~~~
\bra{0}c_{-1} \tilde{c}_{-1}
\tilde{c}_0c_0 \tilde{c}_1 c_1\ket{0}
=i(2\pi)^{26}\delta^{26}(0),
\label{eq:closedIP__0}
\end{eqnarray}
where $\delta^{26}(0)$ is the delta function for momentum conservation.
It is worth noting that, as stated in \cite{Polchinski:1998rq}, the
factor of $i$ is required in the closed string case for hermiticity.
When we factorize a correlation function into matter and ghost sector,
we include $i$ in the ghost correlation function.
Expanding the ghost fields as
\begin{eqnarray}
 c(z)=\sum_{n=-\infty}^\infty c_n z^{-n+1},~~~
 \tilde{c}(\bar{z})=\sum_{n=-\infty}^\infty \tilde{c}_n \bar{z}^{-n+1},
\end{eqnarray}
the inner products (\ref{eq:openIP__0}) and (\ref{eq:closedIP__0}) lead to the
following correlation functions of the ghost sector:
\begin{eqnarray}
 &&
\mathrm{open\ string:}
\nn
&&~~~~~\bra{0}c(z_1)c(z_2)c(z_3) \ket{0}
=(z_1-z_2)(z_1-z_3)(z_2-z_3),
\label{eq:openIP}
\\
&&
\mathrm{closed\ string:}\nn
&&~~~~~
\bra{0}
c(z_1)c(z_2)c(z_3)
\tilde{c}(\bar{z}_1)\tilde{c}(\bar{z}_2)\tilde{c}(\bar{z}_3)
\ket{0}
=-i\,|z_1-z_2|^2|z_1-z_3|^2|z_2-z_3|^2.
\label{eq:closedIP}
\end{eqnarray}
Once again, we observe the presence of the factor $i$ in the closed
string case, which is unconventional. However, we adopt this definition
in order to clarify the correspondence between the one-form $dz$ and the
ghost field $c(z)$. 

When dealing with closed strings, according to this correspondence, the
unintegrated vertex operator should be regarded as $ic\tilde{c}V(z,\bar{z})$.
The expression, which has an unconventional factor of $i$, is naturally
derived from the correspondence between the one-form and the ghost (\ref{eq:dzc}) as
follows:
\begin{eqnarray}
 d^2 z V(z,\bar{z})=i dz\wedge d\bar{z}\,V(z,\bar{z})
&\longrightarrow&
ic\tilde{c}\,V(z,\bar{z}),
\label{eq:fixedclosedvertex}
\end{eqnarray}
where, in addition to the imaginary factor,
we adopt the opposite sign compared to the closed string vertex operator
in \cite{Polchinski:1998rq}.
This is feasible because the overall sign of ghost amplitudes are
chosen to correspond to the positive FP determinant.

We will verify that (\ref{eq:closedIP}) and (\ref{eq:fixedclosedvertex})
lead to the conventional amplitude by examining the expectation value of
closed string vertices. In the case of closed strings, the sphere amplitude
can be related to the correlation function in the
following way:
\begin{eqnarray}
e^{-2\lambda} \Big<\cdots \Big>_{S_2}=iC_{S_2}\,\bra{0}\cdots \ket{0},
\end{eqnarray}
where $e^{-2\lambda}$ represents a contribution related to the Euler characteristic
in \cite{Polchinski:1998rq},
$C_{S_2}$ is a normalization constant for a sphere amplitude and
the factor of $i$ is inserted as discussed in \cite{Polchinski:1998rq}. This factor is
different from the previous $i$ and is needed for the $S$-matrix
calculations in Minkowski space-time.
The amplitude for three closed string vertices $V_i(z_i,\bar{z}_i)\
(i=1,2,3)$ is
\begin{eqnarray}
&&
 g_{\rm c}^3\,e^{-2\lambda}
\Big<
ic\tilde{c}V_1(z_1,\bar{z}_1)\,
ic\tilde{c}V_2(z_2,\bar{z}_2)\,
ic\tilde{c}V_3(z_3,\bar{z}_3)\Big>_{S_2}
\nn
&&=
g_{\rm c}^3iC_{S_2}
\times |z_1-z_2|^2|z_1-z_3|^2|z_2-z_3|^2
\times \bra{0}
V_1(z_1,\bar{z}_1)\,
V_2(z_2,\bar{z}_2)\,
V_3(z_3,\bar{z}_3)\ket{0},
\end{eqnarray}
where the second and third factors correspond to ghost and matter
sector, respectively. It should be noted that, in this case, the ghost
sector is precisely equal to the FP determinant, despite the
use of unconventional definitions. This is because the factor of $i$
appears four times from the prefactor of the inner product and three closed
string vertices, and the combined factor equals $1$ due to $i^4=1$.
Thus, this amplitude becomes equal to the conventional one.

Our convention for the inner product and the closed string vertex
operator may seem unconventional compared to the ordinary ones. However,
our convention is elegant and effective from the perspective of
consistency when dealing with open-closed string amplitudes. Similar
to the case of the sphere, the expectation value on the disk
is related to the correlation function as
\begin{eqnarray}
e^{-\lambda}\Big<\cdots \Big>_{D_2}&=& iC_{D_2}\bra{0}\cdots \ket{0}, 
\end{eqnarray} 
where $C_{D_2}$ is a normalization constant for a disk amplitude.  For
one closed and one open string amplitudes, the following ghost
correlation function appears in the amplitude:
\begin{eqnarray}
 \bra{0}\,ic(z)\tilde{c}(\bar{z})\,c(x)\ket{0},
 \label{eq:corr__icztilczbartilcx}
\end{eqnarray}
where the coordinate $z$ denotes the position of the closed string
vertex on the upper half plane, while $x$ represents the boundary
coordinate of the open string vertex on the real axis.  Using the correlation
 (\ref{eq:openIP}) and the ordinary doubling trick for ghosts,
the correlation (\ref{eq:corr__icztilczbartilcx}) is computed as $ i(z-\bar{z})|z-x|^2$ and then it is
real. Here, it should be noted that our convention for the closed string
vertex operator includes a factor of $i$, which gives rise to a
real-valued ghost correlation function. This is a contrast to the
conventional definition of vertex operator, which does not have $i$ and
leads to an imaginary-valued correlation function. Here, it is noted that
in the conventional case, an extra phase factor must be multiplied to
the ghost correlation function corresponding to the FP
determinant. However, one of the superior points of our convention is
that it does not require such a complex phase.  In the following
section, we will demonstrate that our convention allows for a natural
and straightforward construction of closed string vertex operators with
various ghost numbers, without the need for complex factors that are
required in the conventional approach.

\subsection{A vertex operator with ghost number three}

We begin with a disk amplitude with insertions of closed strings.  We
represent the disk as the upper-half plane $H^+$ and so the CKG is 
$PSL(2,\mathbb{R})$.  Writing a closed string vertex
operator explicitly, the amplitude is expressed as
\begin{eqnarray}
 A=\frac{1}{{\rm vol}PSL(2,\mathbb{R})}\times e^{-\lambda}\,
\,\Big<g_{\rm c}\int_{H^+} d^2 z V(z,\bar{z})
\cdots \Big>_{D_2},
\label{eq:amp1__0}
\end{eqnarray}
where $g_{\rm c}$ is the closed string coupling constant assigned to the
closed string vertex operator as in \cite{Polchinski:1998rq}.
The abbreviation implies insertions
of other vertex operators.
The amplitude is defined by dividing the volume of $PSL(2,\mathbb{R})$.
The transformation $\theta\in PSL(2,\mathbb{R})$ is given by
\begin{eqnarray}
 \theta:~~~z~~\rightarrow~~ z^\theta=\frac{az+b}{cz+d}
\end{eqnarray}
for the real numbers $a,\,b,\,c,\,d$ satisfying $ad-bc=1$.
Conventionally, since $PSL(2,\mathbb{R})$ has three degrees of freedom, the
positions of three open string vertex operators, or one closed and one
open string vertex operators are fixed and multiplied by ghost fields in
the gauge fixing procedure of $PSL(2,\mathbb{R})$.

Let us first consider the case that the positions of one closed and one
open string vertex operators are fixed. The gauge fixing condition is
\begin{eqnarray}
 z^\theta-\hat{z}=0,~~~x^\theta-\hat{x}_0=0,
\end{eqnarray}
where $z$ and $x$ are the positions of the closed and open string
vertex operators and $\hat{z}$ and $\hat{x}_0$ are corresponding fixed
positions, respectively.  To extract the gauge volume under this gauge
fixing condition, let us calculate the FP determinant
$\Delta(\hat{z},\hat{z}^*,\hat{x}_0)$ defined by\footnote{For $z=x+iy$ ($x,y\in \mathbb{R}$), 
the delta function is defined as
\begin{eqnarray}
 \delta^2(z)\equiv \frac{1}{2}\delta(x)\delta(y).
\end{eqnarray}}
\begin{eqnarray}
 \Delta(\hat{z},\hat{z}^*,\hat{x})^{-1}=\int d\mu(\theta)\,
\delta^2({z^\theta-\hat{z}})\delta(x^\theta-\hat{x}_0).
\label{eq:FP1}
\end{eqnarray}
Here, $d\mu(\theta)$ is the gauge invariant measure and,
by using this measure, the gauge volume is formally expressed as
\begin{eqnarray}
 {\rm vol}PSL(2,\mathbb{R})=\int d\mu(\theta).
\label{eq:gaugevol}
\end{eqnarray}
If $\theta$ is parameterized by three positions $x_i\ (i=1,2,3)$ on
the real axis as in the conventional case,
the measure is conventionally represented by
\begin{eqnarray}
 d\mu(\theta)=\frac{dx_1dx_2dx_3}{|x_1-x_2||x_1-x_3||x_2-x_3|},
\label{eq:measurex}
\end{eqnarray}
and this expression fixes the normalization of the measure.
In the standard way, 
the integration of (\ref{eq:FP1}) can be computed
by using the infinitesimal transformation of $PSL(2,\mathbb{R})$:
\begin{eqnarray}
 \Delta(\hat{z},\hat{z}^*,\hat{x}_0)^{-1}&=&\frac{1}{|\hat{z}-\hat{z}^*|
|\hat{z}-\hat{x}_0|^2}.
\label{eq:FP2}
\end{eqnarray}
By following the standard technique, this FP determinant can be
expressed as a correlation function of ghost fields inserted at the
fixed positions of the closed and open string vertices. 

Here, by integrating (\ref{eq:FP1}) and (\ref{eq:FP2}) with respect to
$\hat{x}_0$, we obtain the following equation:
\begin{eqnarray}
\frac{1}{4\pi}\,2|\hat{z}-\hat{z}^*|^2 \,
\,\int d\mu(\theta)\delta^2(z^\theta-\hat{z})=1.
\label{eq:FP3}
\end{eqnarray}
This equation corresponds to a kind of FP formula that applies when
the position of only one closed string is fixed. This is made possible
because the remaining gauge degree of freedom is compact, despite the
fact that $PSL(2,\mathbb{R})$ has three real degrees of freedom.  Inserting
the formula
(\ref{eq:FP3}) into (\ref{eq:amp1__0}) and factorizing out the gauge volume
(\ref{eq:gaugevol}),
we can express the amplitude as
\begin{eqnarray}
 A=ig_{\rm c}\,C_{D_2}\,\bra{0}
\frac{i}{4\pi}(
\partial c-\bar{\partial}\tilde{c})\,ic\tilde{c}
V(\hat{z},\hat{z}^*)\cdots \ket{0}.
\label{eq:vopgh3}
\end{eqnarray}
This expression represents the amplitude with the position 
of only one closed string fixed, where the fixed
vertex operator of the closed string has ghost number three.
It can be observed that this vertex operator is obtained
by multiplying the ghost operator (\ref{eq:dghfac})
with the fixed closed string vertex, 
which includes the factor ``$i$'' as explained above.

This fixed vertex operator corresponds to the state of a linear
combination of the second and third terms of the absolute cohomology 
(\ref{eq:abscohomology}):
\begin{eqnarray}
 V(0,0)\ket{0}\otimes c_0^- \,\tilde{c}_1 c_1\ket{0}.
\label{eq:c0-state}
\end{eqnarray}
Indeed, it is
straightforward to establish the BRST and conformal invariance of the
vertex operator.

It should be noted that the normalization of the vertex operator with
ghost number three can be determined as $1/4\pi$ by the computation of a
finite quantity.  Based on the calculations of \cite{Douglas:1986eu},
which applied to the disk tadpole amplitude, obtaining the normalization
factor involves a type of regularization due to the divergence of a
functional determinant. However, our approach, which uses the
FP technique, circumvents the need for regularization.

This vertex operator discussed above has ghost number three
and could be potentially used to calculate the dilaton tadpole,
 as in \cite{Douglas:1986eu}. However, when the dilaton has zero momentum,
a subtlety arises and a different approach is necessary.
We will address this issue in later sections.

\subsection{A vertex operator with ghost number two}

Next, let us consider the gauge fixing condition in which we fix the
position of one open string vertex and the real part of that of one closed
string vertex:
\begin{eqnarray}
 \mathrm{Re}(z^\theta)-\hat{x}=0,~~~x^\theta-\hat{x}_0=0.
\label{eq:gf2}
\end{eqnarray}
Similarly to the previous case, it is possible to fix only two degrees
of freedom of $PSL(2,\mathbb{R})$ in this situation for the same reason.

To derive the FP formula corresponding to (\ref{eq:gf2}),
we multiply (\ref{eq:FP1}) and (\ref{eq:FP2}) by $|\hat{z}-\hat{z}^*|$:
\begin{eqnarray}
 |\hat{z}-\hat{z}^*|\int d\mu(\theta)
\delta^2({z^\theta-\hat{z}})\delta(x^\theta-\hat{x}_0)
=\frac{1}{|\hat{z}-\hat{x}_0|^2}.
\end{eqnarray}
Then by integrating it with respect to $\hat{y}=\mathrm{Im}\hat{z}$,
we obtain the formula
\begin{eqnarray}
 \int_0^\infty d\hat{y} \,\int d\mu(\theta)\,\frac{2}{\pi}
|\hat{x}-\hat{x}_0||\hat{z}-\hat{z}^*|
\,\delta^2({z^\theta-\hat{z}})\delta(x^\theta-\hat{x}_0)=1.
\end{eqnarray}
Applying the same technique of inserting this formula, we can obtain the
amplitude involving ghost fields. However, it is important to note that
the integration over $\hat{y}$ introduces a discrete symmetry. This residual symmetry 
is $\mathbb{Z}_2$ obtained from combining the inversion
of the real axis and an element of $PSL(2,\mathbb{R})$. As a result, we can
transform the position of the closed string vertex to another point with
a different imaginary part. Therefore, the result obtained from the
straightforward computation must be divided by two for the symmetry.
The resulting amplitude under the gauge fixing
(\ref{eq:gf2}) is expressed as
\begin{eqnarray}
 \mathrm{sgn}(\hat{x}_0-\hat{x})\,
iC_{D_2}\bra{0}
\int_0^\infty d\hat{y}\,\frac{i}{4\pi}
(
\partial c-\bar{\partial}\tilde{c}
)\,
(c+\tilde{c})
V_{\rm c}(\hat{z},\hat{z}^*)
\,cV_{\rm o}(\hat{x}_0)\,\cdots \ket{0},
\label{eq:amp2__0}
\end{eqnarray}
where $V_{\rm c}$ is a $(1,1)$ primary field for closed string, whereas
$V_{\rm o}$ is a primary field for open string with weight $1$.  The overall
sign is given such that the FP determinant is positive.

We can easily confirm that the ghost number two operator produces desired
results by computing an amplitude of one closed and two open string
tachyons:\footnote{The simplest open-closed tachyon amplitude involves a
single open and a single closed string tachyon. However, this amplitude is
rendered zero due to on-shell conditions and momentum conservation.}
\begin{eqnarray}
&&
 g_{\rm c}g_{\rm o}^2\,iC_{D_2}
\bra{0}\int_0^\infty dy_1\,\frac{i}{4\pi}(\partial c-\bar{\partial}\tilde{c})
(c+\tilde{c}) e^{ip_1\cdot X(z_1,\bar{z}_1)}
\,ce^{ip_2\cdot X(x_2)}\,\int_{-\infty}^\infty dx_3 e^{ip_3\cdot X(x_3)}
\ket{0}
\nn
&&=(2\pi)^{26}i\delta(p_1+p_2+p_3)\times 2\pi\,{\rm sgn}(x_{21})\frac{g_{\rm c}}{\alpha^{\prime}},
\label{eq:amp1cl2op}
\end{eqnarray}
where we define as $x_{ij}\equiv x_i-x_j$.
This result agrees with the correct amplitude, differing only by a
sign factor.

Thus, the operator appearing first in the amplitude (\ref{eq:amp2__0}) is
the closed vertex operator with ghost number two. It should be noted
that this operator includes an integration, while the conventional
operator with ghost number two has a fixed position. Similar to the
ghost number three vertex, we observe that this operator also consists
of the conventional vertex (\ref{eq:gh1op}) and the additional ghost
operator (\ref{eq:dghfac}). Furthermore, we will demonstrate that the
operator is invariant under the BRST and conformal transformations,
although the details of these invariances will be discussed in section \ref{sec:descent}.

\subsection{Sphere amplitudes, gauge slices and generic paths}

In this subsection, we will compute sphere amplitudes using the vertex
operators constructed above, in order to verify their effectiveness on
different topological surfaces.

Let us consider the amplitude for three closed string tachyons. First,
we will evaluate using the vertex operator with ghost number three, and
for the total ghost number, we will insert the vertex (\ref{eq:gh1op})
with ghost number one as one of the three tachyons. The amplitude is given
by the correlation function
\begin{eqnarray}
ig_{\rm c}^3C_{S_2}
\bra{0}ic\tilde{c}e^{ip_1\cdot X(z_1,\bar{z}_1)}
\,\frac{i}{4\pi}(\partial c-\bar{\partial}\tilde{c})\,ic\tilde{c}
e^{ip_1\cdot X(z_2,\bar{z}_2)}
\int_{-\infty}^\infty dy_3 (c+\tilde{c})e^{ip_3\cdot X(z_3,\bar{z}_3)}
\ket{0}.
\label{eq:3tamp1}
\end{eqnarray}
We can easily compute this correlation function and the result is
\begin{eqnarray}
\frac{1}{2}\{
\mathrm{sgn}(x_{13})-\mathrm{sgn}(x_{23})
\}\times ig_{\rm c}^3 C_{S_2}(2\pi)^{26}
\delta^{26}(p_1+p_2+p_3).
\end{eqnarray}
This is the correct amplitude up to the sign factor. To obtain the
full amplitude, we simply need to multiply this factor by the
correlation function. Here, it is important to note that there are cases
where the amplitude vanishes, namely when $x_1,\ x_2>x_3$ or $x_1,\
x_2<x_3$.  This seems to reflect that there are cases where our vertex
operators do not fix the residual symmetry, such as $PSL(2,\mathbb{C})$,
correctly. 

By examining the disk amplitude as a simpler example, we illustrate the
relationship between the choice of integration path and the gauge fixing
of residual symmetry. We consider a disk amplitude of one closed and two
open string tachyons. When using the closed string vertex (\ref{eq:gh1op})
with the ghost number one,
the amplitude is proportional to the following expression:
\begin{eqnarray}
 \bra{0}\int_0^\infty dy_1(c+\tilde{c})e^{ip_1\cdot X(z_1,\bar{z}_1)}
ce^{ip_2\cdot X(x_2)}ce^{ip_3\cdot X(x_3)}\ket{0}.
\end{eqnarray}
This correlation function is equivalent to the following integral,
with the factor representing momentum conservation omitted:
\begin{eqnarray}
\int_0^\infty dy_1 8(x_{23})^3\frac{(x_{12}x_{13}-y_1^2)y_1^2}{
\{(x_{12})^2+y_1^2\}^2\{(x_{13})^2+y_1^2\}^2}
&=&
2\pi \{{\rm sgn}(x_{12})+{\rm sgn}(x_{31})\}.
\end{eqnarray}
This result implies that for the amplitude to be non-zero, the
integration path must pass through the two points where open string
tachyons are inserted. This can be understood by considering the
following.  Suppose that we fix first the position of the two open
string vertices for gauge fixing of $PSL(2,\mathbb{R})$. Then, it is seen that the
residual symmetry allows us to move the position of the closed string
vertex along the circle of Apollonius, which is generated by the
points that have the same ratio of distances to the two open string vertex
positions. Therefore, each of the Apollonius circles corresponds
to a gauge orbit of the residual symmetry. Hence, the remaining
integration path along the gauge slice must pass through the two open
string vertices.

Here, it is worth noting that we can express (\ref{eq:3tamp1}) in terms of
the integration along a generic path:
\begin{eqnarray}
ig_{\rm c}^3C_{S_2}
\bra{0}ic\tilde{c}e^{ip_1\cdot X(z_1,\bar{z}_1)}
\,\frac{i}{4\pi}(\partial c-\bar{\partial}\tilde{c})ic\tilde{c}
e^{ip_1\cdot X(z_2,\bar{z}_2)}
\,
i \int_{\partial D} (d\bar{z}_3 c-dz_3 \tilde{c})e^{ip_3\cdot X(z_3,\bar{z}_3)}
\ket{0},
\label{eq:3tamp2}
\end{eqnarray}
where $\partial D$ encloses a domain $D$. It can be shown by a simple
computation that this expression is zero when the domain $D$ 
includes both $z_1$ and $z_2$, or neither $z_1$ nor $z_2$. Otherwise, it
provides the correct amplitude up to a sign, which depends on whether
the domain $D$ includes $z_1$ or $z_2$.  This generalization suggests
that the vertex operator (\ref{eq:gh1op}) is regarded as a one-form
operator. Furthermore, this operator can be rewritten as
\begin{eqnarray}
i
\int_{\partial D} (d\bar{z}_3 c-dz_3 \tilde{c})e^{ip_3\cdot X(z_3,\bar{z}_3)}
=
\int_{D} d^2 z_3\,\bm{\delta}_{\rm B} e^{ip_3\cdot X(z_3,\bar{z}_3)},
\label{eq:dcX}
\end{eqnarray}
where $\bm{\delta}_{\rm B} {\cal O}$ denotes the BRST transformation of
the operator ${\cal O}$. If we were to change the order of the $z_3$-integration and
$\bm{\delta}_{\rm B}$ in this operator, it would become BRST exact,
resulting in a zero amplitude. However, since this operator is
considered as a distribution, the amplitude is given as a non-zero
result. In fact, we can explicitly observe that the correlation function
with the insertion of $\bm{\delta}_{\rm B} e^{ip_3\cdot X}$ contains a
delta function singularity, which leads to a non-zero result after the
$z_3$-integration.

\section{Closed string vertices and descent equations
\label{sec:descent}
}

The equation (\ref{eq:3tamp2}) suggests 
that the vertex operator (\ref{eq:gh1op}) corresponds to the one-form operator,
\begin{eqnarray}
 {\omega_1}^1 = dz\,i\tilde{c}V(z,\bar{z})-d\bar{z}\,icV(z,\bar{z}).
\label{eq:w11}
\end{eqnarray}
Here, we adopt the notation that the subscript of $\omega$ denotes the
rank of differential forms and the superscript indicates the ghost
number of $\omega$. Since $V(z,\bar{z})$ is a $(1,1)$ primary field, and
$d\bar{z}c(z)$ and $dz\tilde{c}(\bar{z})$
have the weight $(-1,-1)$, the form
${\omega_1}^1$ is invariant under conformal transformations.

The BRST transformation of ${\omega_1}^1$
is calculated as
\begin{eqnarray}
 \bm{\delta}_{\rm B}{\omega_1}^1 = d\bar{z}
\bar{\partial}\left(ic\tilde{c} V(z,\bar{z})\right)
+dz\partial \left(ic\tilde{c}V(z,\bar{z})\right),
\label{eq:dBw11}
\end{eqnarray}
which leads to the BRST invariance of the integration of ${\omega_1}^1$
when the surface term vanishes.  Here, we find that the operator on the
right-hand side of (\ref{eq:dBw11}) is a conventional closed string
vertex operator. Therefore, we introduce zero- and two-form operators
that correspond to closed string vertex operators as follows:
\begin{eqnarray}
 {\omega_2}^0&=& i dz\wedge d\bar{z}\,V(z,\bar{z}),
\label{eq:w20}
\\
 {\omega_0}^2&=& i c\tilde{c}V(z,\bar{z}).
\label{eq:w02}
\end{eqnarray}
${\omega_2}^0$ and ${\omega_0}^2$ correspond to conventional integrated
and unintegrated vertex operators, respectively.  Similar to
(\ref{eq:w11}), these operators are also invariant under conformal
transformations.  Let $d$ be the exterior derivative on forms, and
assume that $d$ and $\bm{\delta}_{\rm B}$ commute.  Then, we can find
that these differential form operators satisfy the descent equations:
\begin{eqnarray}
&&
 \bm{\delta}_{\rm B}{\omega_2}^0=-d{\omega_1}^1,
\nn
&&
 \bm{\delta}_{\rm B}{\omega_1}^1=d{\omega_0}^2,
\nn
&&
 \bm{\delta}_{\rm B}{\omega_0}^2=0.
\label{eq:de1}
\end{eqnarray}

In bosonic string theory, these descent equations
were used to study the semi-relative condition for closed strings in
\cite{Becchi:1993pp}.  Additionally, in \cite{Bergman:1994qq}
and \cite{Belopolsky:1995vi}, a momentum zero dilaton state corresponds
to a form that satisfies a part of the equations.  In our case,
(\ref{eq:w11}), (\ref{eq:w20}) and (\ref{eq:w02}) are regarded as
extended forms covering a wider range of quantum operators.

As seen in (\ref{eq:vopgh3}) and (\ref{eq:amp2__0}) in the previous
subsections, we construct unconventional vertex operators with ghost number
three and two. These correspond to the following
differential form operators:
\begin{eqnarray}
 {\omega_1}^2 &=&
\frac{1}{4\pi}(\partial c-\bar{\partial}\tilde{c})
\left(dz\,\tilde{c}V(z,\bar{z})
-d\bar{z}\,cV(z,\bar{z})
\right),
\nn
 {\omega_0}^3 &=&
-\frac{1}{4\pi}(\partial c-\bar{\partial}\tilde{c})c\tilde{c} V(z,\bar{z}).
\end{eqnarray}
These forms are given by multiplying the operator (\ref{eq:dghfac}) to
$-{\omega_1}^1$ and ${\omega_0}^2$. Similarly, ${\omega_2}^1$
is constructed using (\ref{eq:dghfac}) and (\ref{eq:w20}):
\begin{eqnarray}
 {\omega_2}^1&=&
-\frac{1}{4\pi}(\partial c-\bar{\partial}\tilde{c})
dz\wedge d\bar{z}V(z,\bar{z}),
\end{eqnarray}
and these satisfy the descent equations:
\begin{eqnarray}
&&
 \bm{\delta}_{\rm B}{\omega_2}^1=-d{\omega_1}^1,
\nn
&&
 \bm{\delta}_{\rm B}{\omega_1}^2=d{\omega_0}^3,
\nn
&&
 \bm{\delta}_{\rm B}{\omega_0}^3=0.
\label{eq:de2}
\end{eqnarray}
These equations guarantee the BRST invariance of amplitudes that involve
the corresponding vertex operators.

It is evident that ${\omega_0}^3$ shares the same conformal invariance as
${\omega_2}^0$, ${\omega_1}^1$ and ${\omega_0}^2$. However, 
in contrast to these operators,
the operators ${\omega_2}^1$ and ${\omega_1}^2$
do not possess the conformal invariance.
However, despite this lack of invariance,
the amplitudes involving ${\omega_2}^1$ or 
${\omega_1}^2$
remain invariant under conformal
transformations.  This is because these forms are transformed by the
conformal transformation $z'=f(z)$ as follows:
\begin{eqnarray}
 {\omega'_2}^1&=&
{\omega_2}^1+d\left(
-\frac{i}{4\pi}{\cal F}\,{\omega_1}^1
\right)
+\bm{\delta}_{\rm B}\left(
-\frac{i}{4\pi}{\cal F}\,
{\omega_2}^0\right),
\\
 {\omega'_1}^2&=&
{\omega_1}^2+d\left(
-\frac{i}{4\pi}{\cal F}\,{\omega_0}^2\right)
-\bm{\delta}_{\rm B}\left(
-\frac{i}{4\pi}{\cal F}\,
{\omega_1}^1\right),
\end{eqnarray}
where ${\cal F}$ is defined by
\begin{eqnarray}
 {\cal F}=\log\frac{\partial f(z)}{\bar{\partial}\bar{f}(\bar{z})}.
\label{eq:calF}
\end{eqnarray}
The total derivative terms may vanish after integration and the BRST
exact terms become zero in the amplitude.

We can conclude that the forms for closed string vertex operators, which
are constructed by the FP procedure, satisfy the descent
equations. In addition, we have encountered ${\omega_2}^1$ in the
descent equations that is not derived from the FP procedure
discussed in the previous section. Notably, ${\omega_2}^1$ is a two-form
that does not reduce the number of integrations, thus it adds to the
ghost number without appearing to be directly related to gauge-fixing.

Finally, we will check that ${\omega_2}^1$ leads to a correct amplitude. As
an example, we consider the amplitude of a closed string tachyon and two
open string tachyons as follows,
\begin{eqnarray}
 g_{\rm c}g_{\rm o}^2 iC_{D_2}
\bra{0}\int i\,d^2 z
\frac{1}{4\pi}(\partial c-\bar{\partial}\tilde{c})e^{ip\cdot X(z,\bar{z})}
ce^{ip_1\cdot X(x_1)}ce^{ip_2\cdot X(x_2)}\ket{0}.
\end{eqnarray}
Since the two positions of open string vertices are fixed, an additional degree of
freedom must be fixed for the gauge fixing of $PSL(2,\mathbb{R})$.
However, the current expression of the amplitude includes
the integration over the position of the closed string vertex
and does not fix any other degrees of freedom apart from the two positions.
Consequently, this expression of amplitudes involves redundant integrations. 
Taking $z=x+iy$, the amplitude is calculated as
\begin{eqnarray}
&&g_{\rm c}g_{\rm o}^2 C_{D_2}(2\pi)^{26}\delta^{26}(p+p_1+p_2)
\nn
&&
\times
\frac{1}{4\pi} x_{21}|x_{21}|^2
\int_0^\infty dy
\int_{-\infty}^\infty dx 
\frac{32i y^3}{\{(x-x_1)^2+y^2\}^2\{(x-x_2)^2+y^2\}^2}.
\end{eqnarray}
This integration converges although only two positions are fixed and then the resulting amplitude is given by
\begin{eqnarray}
 &&
(2\pi)^{26}i\delta^{26}(p+p_1+p_2)\times 2\pi \,{\rm sgn}(x_{21})\frac{g_{\rm c}}{\alpha^{\prime}},
\end{eqnarray}
which coincides with the correct amplitude except for the sign factor,
similar to (\ref{eq:amp1cl2op}).

\section{Dilaton vertex operators
\label{sec:Dilaton}
}

In this section, we examine dilaton vertex operators with various ghost
numbers, which present a distinct situation in that they do not
straightforwardly satisfy the conformal and BRST invariance as observed
in the other vertices.

\subsection{Dilaton two-forms with ghost number zero
\label{sec:dilaton2forms}}

Let us consider a dilaton vertex operator with ghost number zero:
\begin{eqnarray}
 \partial X\cdot \bar{\partial} X e^{ip\cdot X}~~~(p^2=0).
\label{eq:dilatonvop1}
\end{eqnarray}
It is well-known that this is not a primary field for nonzero momentum. 
When the conformal mode of the world-sheet metric is
taken into account, the dilaton vertex operator incorporates a term
proportional to the scalar curvatures
\cite{Fradkin:1985ys,deAlwis:1985zy,Callan:1986ja}, which is essential
for achieving conformal invariance.  However, in accordance with the
conventional BRST approach that does not incorporate the conformal mode,
we initiate with the dilaton vertex operator (\ref{eq:dilatonvop1}).

To treat the dilaton vertex operator as a primary operator, independently of 
the conformal mode, various expressions can be considered.
First, it is given by using a polarization tensor:\footnote{
We omit the overall normalization factor $2/\sqrt{24}\alpha'$ until the consideration of amplitudes.  
\label{fn:normalization}
}
\begin{eqnarray}
&&
\zeta_{\mu\nu}\,\partial X^\mu \bar{\partial} X^\nu e^{ip\cdot X},~~~
~~~~
\zeta_{\mu\nu}\equiv \eta_{\mu\nu}-p_\mu \bar{p}_\nu
-\bar{p}_\mu p_\nu,
\label{eq:dilatonvop2}
\end{eqnarray}
where $\eta_{\mu\nu}$ is the metric tensor of the Minkowski space and
$\bar{p}_\mu$ is an arbitrary vector satisfying 
$p\cdot \bar{p}=1$ and $\bar{p}\cdot\bar{p}=0$. 
Although this expression represents
a $(1,1)$ primary field, there is a subtlety concerning the momentum zero limit.
Secondly, covariant expression was constructed by incorporating ghost
and anti-ghost fields in \cite{Terao:1987ux}:
\begin{eqnarray}
  \partial X\cdot \bar{\partial} X e^{ip\cdot X}
-\frac{\alpha'}{6}(
cb\bar{\partial}+\tilde{c}\tilde{b}\partial)e^{ip\cdot X},
\label{eq:dilatonvop3}
\end{eqnarray}
which indeed is a $(1,1)$ primary field.

When considering a two-form operator associated with these dilaton
vertex operators, it might be necessary to
consider adopting (\ref{eq:dilatonvop3}) in order to ensure
both conformal and Lorentz invariance.
However, it is easily found that
both two-form operators constructed from (\ref{eq:dilatonvop2}) 
and (\ref{eq:dilatonvop3}) differ from the two-form of 
(\ref{eq:dilatonvop1}) only by a total derivative term:
\begin{eqnarray}
&&d(-\bar{p}\cdot\bar{\partial} Xe^{ip\cdot X}d\bar{z}+\bar{p}\cdot \partial Xe^{ip\cdot X}dz),
\qquad d\left(\frac{i\alpha'}{6}cbe^{ip\cdot X}dz-\frac{i\alpha'}{6}\tilde{c}\tilde{b}e^{ip\cdot X}d\bar{z}\right),
\end{eqnarray}
respectively.
Consequently, we proceed with the following two-form operator constructed from
(\ref{eq:dilatonvop1}):
\begin{eqnarray}
 {\Omega_2}^0=idz\wedge d\bar{z}\,
 \partial X\cdot \bar{\partial} Xe^{ip\cdot X}.
\label{eq:W20}
\end{eqnarray}
In fact, the total derivative term associated with ${\Omega_2}^0$ provides
a trivial contribution in solving the descent equations.
Therefore, (\ref{eq:W20}) alone is sufficient to address the descent equations
and, in particular,
construct a zero-form that follows (\ref{eq:W20}).
It is noted that in the limit of momentum zero, ${\Omega_2}^0$ coincides 
with the two-form that has been investigated in \cite{Belopolsky:1995vi}.

It is worth noting that such a total derivative term becomes significant
when integrating the two-form, as the resulting integration may capture
the global structure of the world-sheet, similar to addressing the soft
dilaton theorem\cite{Bergman:1994qq,Belopolsky:1995vi}.
However, in this paper, we proceed with (\ref{eq:W20})
as we do not delve into further integration of the two-form.

\subsection{Descent equations}

The form operators that satisfy the descent equations, including
${\Omega_2}^0$, can be readily obtained by iteratively applying the BRST
transformation starting from ${\Omega_2}^0$. The resulting operators are given by
\begin{eqnarray}
 {\Omega_1}^1&=& 
dz\left(
i\tilde{c}\partial X\cdot \bar{\partial}X e^{ip\cdot X}
-\frac{i\alpha'}{4}\partial^2ce^{ip\cdot X}
\right)
-d\bar{z}\,\left(
ic\partial X\cdot \bar{\partial}X e^{ip\cdot X}
-\frac{i\alpha'}{4}\bar{\partial}^2\tilde{c}e^{ip\cdot X}
\right),
\label{eq:W11}\\
{\Omega_0}^2&=&
ic\tilde{c}\partial X\cdot \bar{\partial}X e^{ip\cdot X}
-\frac{i\alpha'}{4}(c\partial^2 c-\tilde{c}\bar{\partial}^2 \tilde{c})
e^{ip\cdot X}.
\label{eq:W02}
\end{eqnarray}
These satisfy the following descent equations,
\begin{eqnarray}
&&
 \bm{\delta}_{\rm B}{\Omega_2}^0=-d{\Omega_1}^1,
\nn
&&
 \bm{\delta}_{\rm B}{\Omega_1}^1=d{\Omega_0}^2,
\nn
&&
 \bm{\delta}_{\rm B}{\Omega_0}^2=0.
\end{eqnarray}
Here, ${\Omega_0}^2$ represents a well-known covariant expression of the
dilaton vertex operator. It is worth noting that this expression remains
unchanged regardless of whether we begin with the two-form given by the
vertex operator (\ref{eq:dilatonvop2}) or (\ref{eq:dilatonvop3}).

Let us explore the process of solving the descent equations of
(\ref{eq:de2}) for the dilaton vertex operators. 
First, we consider the BRST transformation of the zero-form which 
is obtained by multiplying ${\Omega_0}^2$ with (\ref{eq:dghfac}):
\begin{eqnarray}
 \bm{\delta}_{\rm B}\left\{
\frac{i}{4\pi}(\partial c-\bar{\partial}\tilde{c}){\Omega_0}^2
\right\}
&=&
-\frac{\alpha'}{8\pi}(c\partial^2 c)(\tilde{c}\bar{\partial}^2 \tilde{c})
e^{ip\cdot X}.
\label{eq:addterm}
\end{eqnarray}
Hence, it becomes necessary to introduce an additional operator to the
zero-form in order to cancel out this term and achieve BRST invariance.
One candidate for the additional operator
is constructed using the vector $\bar{p}_\mu$, which is used
in the polarization tensor in (\ref{eq:dilatonvop2}):
\begin{eqnarray}
&&
 \varDelta {\Omega_0}^3=
-\frac{i}{4\pi}c\tilde{c}(\bar{\partial}^2\tilde{c}\, \bar{p}\cdot
\partial X -\partial^2 c\,\bar{p}\cdot \bar{\partial} X)e^{ip\cdot X}.
\end{eqnarray}
Actually, the BRST transformation of this operator is 
\begin{eqnarray} 
\bm{\delta}_{\rm B} \left(\varDelta {\Omega_0}^3\right)=
+\frac{\alpha'}{8\pi}(c\partial^2 c)(\tilde{c}\bar{\partial}^2 \tilde{c})
e^{ip\cdot X},
\label{eq:deltaBdeltaOmega03}
\end{eqnarray}
and so this operator contributes a cancellation of (\ref{eq:addterm}).
However, this operator exhibits subtleties when considering zero
momentum. Therefore, similar to the case of (\ref{eq:dilatonvop2}), we
choose not to adopt it further. 

Despite our attempt to solve the descent equations,
starting with ${\Omega_0}^2$ multiplied by (\ref{eq:dghfac}),
we were unable to identify
the operators within the range of primary operators and their derivatives
that cancel out the additional term (\ref{eq:addterm}).
Hence,  we encounter difficulties when attempting to directory multiply the
operator (\ref{eq:dghfac}) with ${\Omega_2}^0$, $-{\Omega_1}^1$ or
${\Omega_0}^2$.

If we solely focus on solving the equations using forms with additional
ghost numbers, we can identify the following forms as solutions for the
descent equations:
\begin{eqnarray}
 {\Omega^+_2}^1=(\partial c+\bar{\partial}\tilde{c})
{\Omega_2}^0,
\qquad
 {\Omega^+_1}^2=-(\partial c+\bar{\partial}\tilde{c})
{\Omega_1}^1,
\qquad
 {\Omega^+_0}^3=(\partial c+\bar{\partial}\tilde{c})
{\Omega_0}^2,
\end{eqnarray}
where ${\Omega_2}^0$, ${\Omega_1}^1$ and ${\Omega_0}^2$ are given by
(\ref{eq:W20}), (\ref{eq:W11}) and (\ref{eq:W02}). It is noted that
${\Omega_0^+}^3$ is exactly the BRST invariant vertex operator constructed in
\cite{Terao:1987ux}, which corresponds to the dilaton
state proposed in \cite{Siegel:1984xd}. This operator corresponds to the state obtained by replacing $c_0^-$ with $c_0^+$ in (\ref{eq:c0-state}),
which is included in the absolute cohomology (\ref{eq:abscohomology}).
For operators other than the dilaton, in general, we can solve the
descent equations by multiplying $\partial c+\bar{\partial}\tilde{c}$
with ${\omega_2}^0$, $-{\omega_1}^1$ and ${\omega_0}^2$ as given in
(\ref{eq:w20}), (\ref{eq:w11}) and (\ref{eq:w02}). Furthermore, we can
solve the equations by using both $\partial c+\bar{\partial}\tilde{c}$
and $\partial c-\bar{\partial}\tilde{c}$ simultaneously.

However, 
the cohomology with $c_0^+$ should be
a BRST exact state, as mentioned in the introduction.
Indeed, probably in association with the BRST exactness,
it can be observed that the amplitudes involving the operator
$\partial c+\bar{\partial}\tilde{c}$ yield zero.
Therefore, in this paper, 
we will not further explore the characteristics of the form in terms of $\partial
c+\bar{\partial}\tilde{c}$.

\subsection{A dilaton vertex operator with ghost number three
and ascendants}

In this subsection, we will initiate the construction of dilaton vertex
operators with ghost number three to address the descent equations with
an additional ghost number.

As mentioned in the previous subsection, constructing a ghost number
three primary operator for the dilaton, composed solely of conventional
primary operators, appears to be an impossible task.  So, we will explore
such an operator within the extended space of operators including
the string coordinates $X^\mu$.  It is well-known that the operator
$X^\mu$ is not considered primary and so it is typically avoided when
incorporating it into the framework of conformal field theory.  However,
$X^\mu$ is originally incorporated as an operator in the first quantized
string theory. Moreover, the inclusion of $X^\mu$ offers intriguing
insights into the dilaton theorem \cite{Belopolsky:1995vi}, and the BRST
cohomology is well-defined when the zero-mode of $X^\mu$ is included
\cite{Astashkevich:1995cv}. Here, we will actively use $X^\mu$ in order to
construct the dilaton vertex operator.

From (\ref{eq:addterm}), we have only to construct an additional term
$\varDelta {\Omega_0}^3$ to cancel this anomalous term. To obtain such an
operator, we first consider the following operators
\begin{eqnarray}
 C=X\cdot \partial X e^{ip\cdot X},~~~
 \tilde{C}=X\cdot \bar{\partial} X e^{ip\cdot X}~~~
(\,p^2=0\,).
\end{eqnarray}
We can find the OPEs of $C(z,\bar{z})$ with energy momentum tensors:
\begin{eqnarray}
 T(y)C(z,\bar{z})&\sim&
\frac{1}{(y-z)^3}\left(
-13\alpha' e^{ip\cdot X}(z,\bar{z})
-\frac{\alpha'}{2}ip\cdot X e^{ip\cdot X}(z,\bar{z})\right)
\label{eq:TC}
\nn
&&
+\frac{1}{(y-z)^2}\left(-\frac{\alpha'}{2}\partial e^{ip\cdot X}(z,\bar{z})
+C(z,\bar{z})\right)
+\frac{1}{y-z}\partial C(z,\bar{z}),
\\
 \tilde{T}(\bar{y})C(z,\bar{z})&\sim&
\frac{1}{(\bar{y}-\bar{z})^2}\left(-\frac{\alpha'}{2}
\partial e^{ip\cdot X}(z,\bar{z})\right)
+\frac{1}{\bar{y}-\bar{z}}\bar{\partial} C(z,\bar{z}),
\label{eq:tildeTC}
\end{eqnarray}
and we can provide analogous OPEs for $\tilde{C}(z,\bar{z})$.
These OPEs result in the following BRST transformations of $C$ and $\tilde{C}$:
\begin{eqnarray}
 \bm{\delta}_{\rm B} C&=&\partial^2 c
\left(
-\frac{13\alpha' }{2}e^{ip\cdot X}
-\frac{\alpha'}{4}ip\cdot X e^{ip\cdot X}
\right)
+\partial c
\left(-\frac{\alpha'}{2}\partial e^{ip\cdot X}\right)
+\partial \left(cC\right)
\nn
&&
+\bar{\partial}\tilde{c}
\left(-\frac{\alpha'}{2}
\partial e^{ip\cdot X}\right)+\tilde{c}\bar{\partial}
C,
\label{eq:deltaBC}
\\
 \bm{\delta}_{\rm B} \tilde{C}
&=&\bar{\partial}^2 \tilde{c}
\left(
-\frac{13\alpha' }{2}e^{ip\cdot X}
-\frac{\alpha'}{4}ip\cdot X e^{ip\cdot X}
\right)
+\bar{\partial} \tilde{c}
\left(-\frac{\alpha'}{2}\bar{\partial} e^{ip\cdot X}\right)
+\bar{\partial} \left(\tilde{c}\tilde{C}\right)
\nn
&&
+\partial c
\left(-\frac{\alpha'}{2}
\bar{\partial} e^{ip\cdot X}\right)+c \partial
\tilde{C}.
\label{eq:deltaBCtilde}
\end{eqnarray}
It is noted that these BRST transformations do not close within the set
of prepared operators, and they additionally include 
$p\cdot X e^{ip\cdot
X}$, $\partial e^{ip\cdot X}$ and $\bar{\partial}e^{ip\cdot X}$.

Then, we extend the operators $C$ and $\tilde{C}$ by multiplying a
series of $p\cdot X$:
\begin{eqnarray}
 C(h)\equiv h(p\cdot X)C,~~~
 \tilde{C}(h)\equiv h(p\cdot X)\tilde{C},
\end{eqnarray}
where $h(x)$ is a function defined by the series
\begin{eqnarray}
 h(x)=\sum_{k=0}^\infty h_k\,x^k.
\end{eqnarray}
In addition, we define the operator for the function $h(x)$ as
\begin{eqnarray}
 e(h)\equiv h(p\cdot X)e^{ip\cdot X}.
\end{eqnarray}
The point is that these operators can be written as
\begin{eqnarray}
 C(h)&=&\left.
h\left(-i\hat{p}_\mu \frac{\partial}{\partial p_\mu}\right)C
\,\right|_{\hat{p}=p},
\nn
 \tilde{C}(h)&=&\left.
h\left(-i\hat{p}_\mu \frac{\partial}{\partial p_\mu}\right)\tilde{C}
\,\right|_{\hat{p}=p},
\nn
 e(h)&=&\left.
h\left(-i\hat{p}_\mu \frac{\partial}{\partial p_\mu}\right)e^{ip\cdot X}
\,\right|_{\hat{p}=p}.
\end{eqnarray}
Moreover, we find the equation
\begin{eqnarray}
\left.
 h\left(-i\hat{p}_\mu \frac{\partial}{\partial p_\mu}\right)p\cdot X\,e^{ip\cdot X}
\,\right|_{\hat{p}=p}
=\hat{h}(p\cdot X)e^{ip\cdot X}=e(\hat{h}),
\end{eqnarray}
where $\hat{h}(x)=xh(x)-ix h'(x)$.
Consequently, by applying
$h(-i\hat{p}_\mu \partial/\partial p_\mu)$ to
both sides of equations (\ref{eq:deltaBC}) and (\ref{eq:deltaBCtilde}),
and subsequently replacing $\hat{p}$ with $p$,
we can derive the BRST transformation of $C(h)$ and $\tilde{C}(h)$
as shown below:
\begin{eqnarray}
 \bm{\delta}_{\rm B} C(h)&=&\partial^2 c
\left(
-\frac{13\alpha' }{2}e(h)
-\frac{i\alpha'}{4}e(\hat{h})
\right)
+\partial c
\left(-\frac{\alpha'}{2}\partial e(h)\right)
+\partial \left(cC(h)\right)
\nn
&&
+\bar{\partial}\tilde{c}
\left(-\frac{\alpha'}{2}
\partial e(h)\right)+\tilde{c}\bar{\partial}
C(h),
\label{eq:deltaBCh}
\\
 \bm{\delta}_{\rm B} \tilde{C}(h)
&=&\bar{\partial}^2 \tilde{c}
\left(
-\frac{13\alpha'}{2}e(h)
-\frac{i\alpha'}{4}e(\hat{h})
\right)
+\bar{\partial} \tilde{c}
\left(-\frac{\alpha'}{2}\bar{\partial} e(h)\right)
+\bar{\partial} \left(\tilde{c}\tilde{C}(h)\right)
\nn
&&
+\partial c
\left(-\frac{\alpha'}{2}
\bar{\partial} e(h)\right)+c \partial
\tilde{C}(h).
\label{eq:deltaBCtildeh}
\end{eqnarray}
Similarly, we can obtain
\begin{eqnarray}
 \bm{\delta}_{\rm B}e(h)
=c\partial e(h)+\tilde{c}\bar{\partial} e(h).
\label{eq:deltaBeh}
\end{eqnarray}
By using (\ref{eq:deltaBCh}), (\ref{eq:deltaBCtildeh}) and
(\ref{eq:deltaBeh}),  we can find the following equation
\begin{eqnarray}
&&
 \bm{\delta}_{\rm B}\left\{
\tilde{c}\bar{\partial}^2 \tilde{c} c C(h)
+c\partial^2 c \tilde{c}\tilde{C}(h)
-\frac{\alpha'}{2}(\partial c+\bar{\partial}\tilde{c})
(c\partial^2 c+\tilde{c}\bar{\partial}^2
\tilde{c})e(h)\right\}
\nn
&&=
\frac{\alpha'}{2}
(c\partial^2 c)(\tilde{c}\bar{\partial}^2 \tilde{c})
e(24h+i\hat{h}).
\label{eq:deltaBCeh}
\end{eqnarray}

Therefore, from (\ref{eq:deltaBCeh}), we can construct the supplementary
 term $\varDelta 
{\Omega_0}^3$ by finding the function $h$ which satisfies the equation:
\begin{eqnarray}
 24h(x)+i\hat{h}(x)=\frac{1}{4\pi}.
\end{eqnarray}
Finally, we conclude that, by using the function
\begin{eqnarray}
 h(x)=\frac{1}{4\pi}\sum_{k=0}^\infty \frac{23!}{(24+k)!}(-ix)^k,
\end{eqnarray}
$\varDelta {\Omega_0}^3$ is given by
\begin{eqnarray}
\varDelta {\Omega_0}^3= \tilde{c}\bar{\partial}^2 \tilde{c} c C(h)
+c\partial^2 c \tilde{c}\tilde{C}(h)
-\frac{\alpha'}{2}(\partial c+\bar{\partial}\tilde{c})
(c\partial^2 c+\tilde{c}\bar{\partial}^2
\tilde{c})e(h),
\end{eqnarray}
and this operator satisfies (\ref{eq:deltaBdeltaOmega03}).
As a result, the ghost number three vertex operator for dilaton is given by
\begin{eqnarray}
 {\Omega_0}^3=
-\frac{1}{4\pi}(\partial c-\bar{\partial}\tilde{c})
c\tilde{c} \partial X\cdot \bar{\partial}X e^{ip\cdot X}
+\frac{\alpha'}{16\pi}(\partial c-\bar{\partial}\tilde{c})
(c\partial^2 c-\tilde{c}\bar{\partial}^2 \tilde{c})e^{ip\cdot X}
+\varDelta {\Omega_0}^3,
\label{eq:W03}
\end{eqnarray}
and, by construction, this is invariant under the BRST transformation.

From the expression (\ref{eq:W03}), 
we can find ${\Omega_1}^2$ and ${\Omega_2}^1$
satisfying the descent equations including ${\Omega_0}^3$:
\begin{eqnarray}
 \bm{\delta}_{\rm B}{\Omega_2}^1&=&-d{\Omega_1}^2,
\nn
 \bm{\delta}_{\rm B}{\Omega_1}^2&=&d{\Omega_0}^3,
\nn
 \bm{\delta}_{\rm B}{\Omega_0}^3&=&0,
\end{eqnarray}
where ${\Omega_2}^1$ and ${\Omega_1}^2$ are given by
\begin{eqnarray}
 {\Omega_2}^1 &=& dz\wedge d\bar{z}
\left\{
-\frac{1}{4\pi}(
\partial c-\bar{\partial}\tilde{c})\partial X\cdot \bar{\partial X}e^{ip\cdot X}
+\bar{\partial}^2 \tilde{c}C(h)-\partial^2c \tilde{C}(h)\right\},
\label{eq:W21}
\\
{\Omega_1}^2&=&
d\bar{z}\left\{
-\frac{1}{4\pi}(\partial c-\bar{\partial}\tilde{c})
\left(
c \partial X\cdot \bar{\partial} X-\frac{\alpha'}{4}\bar{\partial}^2
\tilde{c}\right)e^{ip\cdot X}
\right.
\nn
&& \left.
+c\partial^2 c \tilde{C}(h)
+\bar{\partial}^2 \tilde{c} c C(h)
+\frac{\alpha'}{2}
(\partial c+\bar{\partial}\tilde{c})\bar{\partial}^2\tilde{c}\,e(h)\right\}
\nn
&&
+dz\left\{
\frac{1}{4\pi}(\partial c-\bar{\partial}\tilde{c})
\left(
\tilde{c} \partial X\cdot \bar{\partial} X-\frac{\alpha'}{4}\partial^2
c\right)e^{ip\cdot X}
\right.
\nn
&& \left.
+\tilde{c}\bar{\partial}^2 \tilde{c} C(h)
+\partial^2 c\tilde{c}  \tilde{C}(h)+\frac{\alpha'}{2}
(\partial c+\bar{\partial}\tilde{c})\partial^2c\,e(h)\right\}.
\label{eq:W12}
\end{eqnarray}

\subsection{Conformal transformations for the form of dilaton}

First, we will confirm the conformal transformation property of the forms (\ref{eq:W20}),
(\ref{eq:W11}) and (\ref{eq:W02}). 
Under the conformal transformation $z'=f(z)$, we find
that the operator (\ref{eq:dilatonvop1}) is transformed as
\begin{eqnarray}
\partial X'\cdot \bar{\partial} X' e^{ip\cdot X'}(z',\bar{z}')
&=&
 |f'(z)|^{-2}\Big\{
\partial X\cdot \bar{\partial} X e^{ip\cdot X}(z,\bar{z})
\nn
&&
+\frac{i\alpha'}{4}\frac{f''(z)}{f'(z)}p\cdot 
\bar{\partial}X e^{ip\cdot X}(z,\bar{z})
+\frac{i\alpha'}{4}\overline{\frac{f''(z)}{f'(z)}}p\cdot 
\partial X e^{ip\cdot X}(z,\bar{z})\Big\},
\label{eq:cmap1}
\end{eqnarray}
where the second and third terms in the right-hand side correspond to
non-tensor terms. Therefore, the dilaton forms, ${\Omega_2}^0$,
${\Omega_1}^1$ and ${\Omega_0}^2$, do not exhibit simple conformal
invariance.

From (\ref{eq:cmap1}), it follows that, under the
conformal transformation $z'=f(z)$, 
these forms are transformed as
\begin{eqnarray}
 {\Omega'_2}^0 &=&{\Omega_2}^0+d\left(
\frac{i\alpha'}{4}\,{\cal F}
\,de^{ip\cdot X(z,\bar{z})}\right),
\nn
 {\Omega'_1}^1&=&
{\Omega_1}^1+d\left\{
-\frac{i\alpha'}{4}(\partial {\cal F}c+\bar{\partial}{\cal F}\tilde{c})e^{ip\cdot X}\right\}
+\bm{\delta}_{\rm B}\left\{
\frac{i\alpha'}{4}(d{\cal F})\,e^{ip\cdot X}\right\},
\nn
{\Omega'_0}^2&=&{\Omega_0}^2
+\bm{\delta}_{\rm B}\left\{
-\frac{i\alpha'}{4}(\partial {\cal F}c+\bar{\partial}{\cal F}\tilde{c})
e^{ip\cdot X}\right\},
\end{eqnarray}
where ${\cal F}$ is defined by (\ref{eq:calF}), resulting from the
conformal map $z'=f(z)$.  Thus, the integration of ${\Omega_2}^0$
satisfies conformal invariance similar to ${\omega_2}^1$ and
${\omega_1}^2$ in the previous section, and the extra terms vanish
in amplitudes due to integration of forms and BRST invariance.

For the forms (\ref{eq:W03}), (\ref{eq:W12}) and (\ref{eq:W21}), we have
to deal with the conformal transformation $C(h)$ and $\tilde{C}(h)$.
From (\ref{eq:TC}) and (\ref{eq:tildeTC}), we obtain the following OPEs:
\begin{eqnarray}
 T(y)C(h)(z,\bar{z})&\sim&
\frac{1}{(y-z)^3}
\frac{-\alpha' }{2}e(26h+i\hat{h})(z,\bar{z})
\nn
&&
+\frac{1}{(y-z)^2}\left(\frac{-\alpha'}{2}\partial e(h)(z,\bar{z})
+C(h)(z,\bar{z})\right)
+\frac{1}{y-z}\partial C(h)(z,\bar{z}),
\label{eq:TCh}
\\
 \tilde{T}(\bar{y})C(h)(z,\bar{z})&\sim&
\frac{1}{(\bar{y}-\bar{z})^2}\frac{-\alpha'}{2}
\partial e(h)(z,\bar{z})
+\frac{1}{\bar{y}-\bar{z}}\bar{\partial} C(h)(z,\bar{z}).
\label{eq:tildeTCh}
\end{eqnarray}
Similar OPEs hold for $\tilde{C}(h)$. Using
these OPEs, we can deduce the transformation rules of 
$C(h)$ and $\tilde{C}(h)$
under the conformal mapping $z'=f(z)$:
\begin{eqnarray}
 \big(C(h)\big)'&=& (f'(z))^{-1}
\left(C(h)+\frac{\alpha'}{4}\partial {\cal F}\,e(26 h+i\hat{h})
+\frac{\alpha'}{2}\log|f'(z)|^2\,\partial e(h)\right),
\label{eq:cmap2}
\\
 \big(\tilde{C}(h)\big)'&=& (\overline{f'(z)})^{-1}
\left(\tilde{C}(h)+\frac{\alpha'}{4}\bar{\partial}
 {\cal F}\,e(26 h+i\hat{h})
+\frac{\alpha'}{2}\log|f'(z)|^2\,\bar{\partial} e(h)\right).
\label{eq:cmap3}
\end{eqnarray}
Thus, the operators $C(h)$ and $\tilde{C}(h)$ are transformed as
non-tensor fields.  It is noted that, for arbitrary $h$, $e(h)$ is
transformed in the same way as a $(0,0)$ primary field, though it is not
a well-behaved primary field due to $X^\mu$ in $e(h)$.

Using (\ref{eq:cmap1}), (\ref{eq:cmap2}) and (\ref{eq:cmap3}), we can
find that, for the conformal mapping $z'=f(z)$, 
the forms ${\Omega_2}^1$, ${\Omega_1}^2$ and ${\Omega_0}^3$
are transformed as
\begin{eqnarray}
 {\Omega'_2}^1 &=& {\Omega_2}^1+d{\Phi_1}^1
+\bm{\delta}_{\rm B}{\Psi_2}^0,
\\
{\Omega'_1}^2
&=&{\Omega_1}^2+d{\Phi_0}^2-\bm{\delta}_{\rm B}{\Phi_1}^1,
\\
{\Omega'_0}^3
&=&{\Omega_0}^3+\bm{\delta}_{\rm B}{\Psi_0}^2,
\label{eq:cmapW03}
\end{eqnarray}
where ${\Phi_1}^1$, ${\Psi_2}^0$, ${\Phi_0}^2$ and ${\Psi_0}^2$ are given by
\begin{eqnarray}
 {\Phi_1}^1&=& -\frac{i}{4\pi}{\cal F}{\Omega_1}^1
+(d{\cal F})(\partial c-\bar{\partial}\tilde{c}+\partial {\cal F} c
+{\bar \partial}{\cal F}\tilde{c})\frac{\alpha'}{16\pi}e^{ip\cdot X}
\nn
&&
+\frac{\alpha'}{2}\log|\partial f(z)|^2
\left\{\bar{\partial}(\bar{\partial}\tilde{c}-\bar{\partial}{\cal F}\tilde{c})
e(h)d\bar{z}
+\partial(\partial c+\partial {\cal F}c)e(h)dz\right\}
+\bar{\partial}{\cal F}\tilde{c}C(h)dz-\partial{\cal F}c\tilde{C}(h)d\bar{z}
\nn
&&
+\bar{\partial}{\cal F}\left(
\frac{\alpha'}{2}(\partial c+\bar{\partial}\tilde{c})e(h)
-cC(h)\right)d\bar{z}
-\partial{\cal F}\left(
\frac{\alpha'}{2}(\partial c+\bar{\partial}\tilde{c})e(h)
-\tilde{c}\tilde{C}(h)\right)dz,
\\
{\Psi_2}^0
&=&
-\frac{i}{4\pi}{\cal F}{\Omega_2}^0+dz\wedge d\bar{z}
\left(\bar{\partial}{\cal F} C(h)+\partial {\cal F}\tilde{C}(h)\right),
\\
{\Phi_0}^2&=&
-\frac{i}{4\pi}{\cal F}{\Omega_0}^2-(\partial c-\bar{\partial}\tilde{c})
\frac{\alpha'}{16\pi}(\partial {\cal F}c+\bar{\partial}{\cal F}\tilde{c})
e^{ip\cdot X}
-\tilde{c}c\left(\bar{\partial}{\cal F} C(h)
+\partial {\cal F} \tilde{C}(h)\right)
+\frac{\alpha'}{2}(\partial c+\bar{\partial}\tilde{c})e(h)
\nn
&&
+\frac{\alpha'}{2}\log|\partial f|^2(\tilde{c}\bar{\partial}^2
\tilde{c}+c\partial^2 c+\partial{\cal F}c\partial c-\bar{\partial}{\cal F}
\tilde{c}\bar{\partial}\tilde{c})e(h),
\\
{\Psi_0}^2&=&\frac{i}{4\pi}{\cal F}{\Omega_0}^2+{\Phi_0}^2.
\end{eqnarray}
Hence, we have found that, under conformal transformation, all forms
associated with dilaton are accompanied by a total derivative term and
 a BRST exact term.

\section{Dilaton tadpole amplitude in the BRST formalism
\label{sec:Dilatontadpole}
}

Now that we have obtained the forms with various ghost numbers
corresponding to closed string vertex operators, we can calculate
various amplitudes by using these forms in the BRST formalism.  In this
section, even among them, we will focus on dilaton tadpole amplitude,
which has been so far scarcely examined in the BRST formalism.  The
dilaton tadpole is computed from the amplitude with an insertion of a
single dilaton vertex in an upper-half plane. Therefore, it is believed
that this tadpole amplitude can be calculated using ${\Omega_0}^3$ of
(\ref{eq:W03}), and
the resulting amplitude will be examined to verify the well-known
coefficient $(26+2)/(26-2)$ discussed in \cite{Liu:1987nz}.

As in (\ref{eq:cmapW03}), through conformal transformations,
${\Omega_0}^3$ is added by the BRST exact term, which might not
contribute to scattering amplitudes. However, since ${\Omega_0}^3$
includes the operator $X\cdot \partial X$ through $C(h)$ and so on,
${\Omega_0}^3$ is not a true primary field. Because amplitudes
containing this operator do not typically disappear, even if they are
BRST exact,\footnote{This point is addressed in \cite{Belopolsky:1995vi}}
 we must examine the specific surface on which ${\Omega_0}^3$
is defined.

According to the state-operator isomorphism, the operator ${\Omega_0}^3$
should correspond to the state of dilaton with ghost number three.\footnote{
Actually, ${\Omega_0}^2$ (\ref{eq:W02}) is
BRST equivalent to the operator made of (\ref{eq:dilatonvop2}) for
$p^{\mu}\ne 0$ in the ghost number two sector, namely
$\Omega_0^{~2}=ic\tilde{c}\zeta_{\mu\nu}\partial
X^{\mu}\bar{\partial}X^{\nu}e^{ip\cdot X}+{\bm\delta}_{\rm
B}(-c\bar{p}\cdot \partial X e^{ip\cdot X}+\tilde{c}\bar{p}\cdot
\bar{\partial}X e^{ip\cdot X})$.  Similarly,
$\Omega_0^{~3}=\frac{i}{4\pi}(\partial c-\bar{\partial}\tilde
c)\Omega_0^{~2}+\varDelta \Omega_0^{~3}$ can be expressed as 
a sum of three terms,
comprising the dilaton vertex operator with ghost number three
constructed from (\ref{eq:dilatonvop2}):
$-\frac{1}{4\pi}(\partial c-\bar{\partial}\tilde{c})c\tilde{c}\zeta_{\mu\nu}\partial X^{\mu}\bar{\partial}X^{\nu}e^{ip\cdot X}$, 
a BRST exact operator:
${\bm \delta}_{\rm B}\{(\partial c+\bar{\partial}\tilde{c})(cC(h)+\tilde{c}\tilde{C}(h))
-\frac{i}{2\pi}(c\partial c\bar{p}\cdot\partial Xe^{ip\cdot X}
+\tilde{c}\bar{\partial}\tilde{c}\bar{p}\cdot\bar{\partial}Xe^{ip\cdot X})
\}$ and an operator with the factor corresponding to $c_0^+$:
$(\partial c+\bar{\partial}\tilde{c})c\tilde{c}\{
\partial(\tilde{C}(h)+\frac{i}{4\pi}\bar{p}\cdot \bar{\partial}Xe^{ip\cdot X})
-\bar{\partial}(C(h)+\frac{i}{4\pi}\bar{p}\cdot \partial Xe^{ip\cdot X})
\}
$.
Then $\Omega_0^{~3}$ gives a BRST closed state ${\Omega_0}^3(0,0)\ket{0}$,
which includes the zero mode of $X^\mu$.
In \cite{Belopolsky:1995vi,Astashkevich:1995cv}, it is shown that the
extended BRST complex including the zero modes is well-defined.
This state should be equivalent to the dilaton state with ghost number three
since the state including $c_0^+$ should be BRST exact, although
it remains to be proved for ${\Omega_0}^3(0,0)\ket{0}$.}
In this isomorphism, the operator is initially placed at
the origin of a unit disk, and thus, ${\Omega_0}^3$ is also considered
to be situated in the same manner at the outset.

As a result, to obtain the dilaton tadpole, we need to compute the
correlation function of ${\Omega_0}^3$ at the origin of the unit disk.
To evaluate this correlator, we initially transform it into a correlator
on an upper-half plane using the following conformal transformation:
\begin{eqnarray}
 z=g(z')=r_0 \frac{e^{i(\theta_1-\theta_0)}z'+e^{i\theta_0}}{
e^{i\theta_1}z'+1}~~~
(r_0>0,~~\theta_0,\theta_1\in \mathbb{R},~\sin\theta_0>0).
\end{eqnarray}
This mapping is a general expression that provides a one-to-one
correspondence between the unit disk $(z')$ and the upper-half plane
$(z)$, and it relocates the origin of the unit disk to $z_0=r_0
e^{i\theta_0}$ within the upper-half plane $(\,z_0=g(0)\,)$.
Then, the inverse mapping is given by
\begin{eqnarray}
 z'=f(z)=\frac{e^{i(\theta_0-\theta_1)} z -r_0
e^{i(2\theta_0-\theta_1)}}{-e^{i\theta_0}z+r_0}.
\label{eq:zmap}
\end{eqnarray}

Hence, based on (\ref{eq:W03}) and (\ref{eq:cmapW03}),
the dilaton tadpole amplitude
should be determined by computing the following expression:
\begin{eqnarray}
 {\cal A}&=& i\frac{2}{\sqrt{24}\alpha'}g_{\rm c}
C_{D_2}\Big\{
\bra{0}\frac{-1}{4\pi}(\partial c-\bar{\partial}\tilde{c})
c\tilde{c}\partial X\cdot \bar{\partial}X e^{ip\cdot X}(z_0,\bar{z}_0)
\ket{0}
\nn
&&
+\frac{\alpha'}{16\pi}\bra{0}
(\partial c-\bar{\partial}\tilde{c})(c\partial^2 c
-\tilde{c}\bar{\partial}^2 \tilde{c})e^{ip\cdot X}(z_0,\bar{z}_0)\ket{0}
\nn
&&
+\bra{0}\varDelta {\Omega_0}^3(z_0,\bar{z}_0)\ket{0}
+\bra{0}\bm{\delta}_{\rm B}
{\Psi_0}^2\ket{0}(z_0,\bar{z}_0)\Big\},
\label{eq:Adilaton}
\end{eqnarray}
where the normalization factor $2/\sqrt{24}\alpha'$, which is mentioned in footnote \ref{fn:normalization},
is considered, and all correlations are
defined on the upper-half plane using the conventional decoupling trick.
It is worth emphasizing once more that the fourth term does not vanish
even though it is BRST exact,
and the function $f(z)$ contained within
${\Psi_0}^2$ is given by (\ref{eq:zmap}).

The calculation of the first term is straightforward and yields:\footnote{According to \cite{Polchinski:1998rq},
\begin{eqnarray}
 g_{\rm c}C_{D_2}=\frac{g_{\rm c}}{\alpha' g_{\rm o}^2}
=\frac{\pi\sqrt{\pi}}{16}
(4\pi^2 \alpha')^{-7}.
\end{eqnarray}}
\begin{eqnarray}
 i(2\pi)^{26}\delta^{26}(p)\,\frac{2}{\sqrt{24}\alpha'}
\frac{\pi\sqrt{\pi}}{16}
(4\pi^2 \alpha')^{-7}\,
\frac{1}{4\pi}26\alpha'=
i(2\pi)^{26}\delta^{26}(p)\,\frac{26}{2\sqrt{24}}\,
\frac{\sqrt{\pi}}{16}\,
(4\pi^2 \alpha')^{-7},
\label{eq:amp1}
\end{eqnarray}
and the second term evaluates to zero.

To evaluate the third term, we need to consider the correlation
functions of $C(h),\ \tilde{C}(h)$ and $e(h)$, which explicitly involve
$X^\mu$ operators. As discussed in \cite{Belopolsky:1995vi}, such a
correlator should be evaluated using
$X^\mu(z,\bar{z})=-i\frac{\partial}{\partial p_\mu}\exp(ip\cdot
X(z,\bar{z}))\big|_{p=0}$, and so,
$C(h)$ and $\tilde{C}(h)$ should be replaced by
\begin{eqnarray}
 C(h)&=&-i\frac{\partial}{\partial p^\nu}\,h\left(
-i\hat{p}_\mu\frac{\partial}{\partial p_\mu}\right)
\partial X^\nu e^{ip\cdot X}\Big|_{\hat{p}=p,\ p\cdot p=0},
\nn
 \tilde{C}(h)&=&-i\frac{\partial}{\partial p^\nu}\,h\left(
-i\hat{p}_\mu\frac{\partial}{\partial p_\mu}\right)
\bar{\partial} X^\nu e^{ip\cdot X}\Big|_{\hat{p}=p,\ p\cdot p=0}.
\label{eq:Chdef}
\end{eqnarray}
Consequently, the correlators involving $C(h)$ and $\tilde{C}(h)$ in the third term
are found to be zero.
Similarly, the correlator for $e(h)$ is evaluated as
\begin{eqnarray}
 \bra{0}e(h)\ket{0}&=&\frac{1}{4\pi}
\sum_{k=0}^\infty \frac{(23)!}{(24+k)!}(-1)^k
p_{\mu_1}p_{\mu_2}\cdots p_{\mu_k}
\frac{\partial}{\partial p_{\mu_1}}\frac{\partial}{\partial p_{\mu_2}}
\cdots \frac{\partial}{\partial p_{\mu_k}}
(2\pi)^{26}\delta^{26}(p).
\label{eq:ehcal}
\end{eqnarray}
Through ``partial integration'', this can be calculated as the product of 
the delta function and ``a numerical factor''. Therefore, 
the third term of (\ref{eq:Adilaton}) can be expressed as
\begin{eqnarray}
 i(2\pi)^{26}\delta^{26}(p)\,
\frac{1}{2\sqrt{24}}\frac{\sqrt{\pi}}{16}(4\pi^2\alpha')^{-7}
\,\left(-4\sum_{k=0}^\infty \frac{25+k}{25\cdot 24}\right).
\label{eq:amp2}
\end{eqnarray}
It is important to emphasize that the numerical factor diverges and 
requires regularization.
However, it should be noted that this divergence arises 
from the correlator of $e(h)$, and in any case, we will continue with the calculation.

Regarding the fourth term, we need to evaluate the correlator of
\begin{eqnarray}
 \bm{\delta}_{\rm B}{\Psi_0}^2
&=&
\partial {\cal F}\left\{
\tilde{c}c\partial c\tilde{C}(h)
-\frac{\alpha'}{2}\bar{\partial}
\tilde{c}c\partial c\Bigl(e(h)+\frac{e^{ip\cdot X}}{8\pi}\Bigr)\right\}
-
\bar{\partial} {\cal F}\left\{
\tilde{c}\bar{\partial}\tilde{c} c C(h)
-\frac{\alpha'}{2}
\tilde{c}\bar{\partial}\tilde{c}
\partial c
\Bigl(e(h)+\frac{e^{ip\cdot X}}{8\pi}\Bigr)\right\}
\nn
&&
+\frac{\alpha'}{2}\log|\partial f|^2\left\{
(\tilde{c}\bar{\partial}^2 \tilde{c}-\bar{\partial}{\cal F}
\tilde{c}\bar{\partial}\tilde{c})c\partial e(h)
+\tilde{c}(c\partial^2 c+\partial {\cal F}c\partial c)\bar{\partial}e(h)\right\}
\nn
&&
+\frac{\alpha'}{16\pi}(\partial c-\bar{\partial}\tilde{c})
\tilde{c}c\left(\partial {\cal F}\bar{\partial}e^{ip\cdot X}
-\bar{\partial}{\cal F}\partial e^{ip\cdot X}\right).
\label{eq:delBPsi20}
\end{eqnarray}
It is easily found that the correlators of $C(h)$, $\partial e(h)$,
$\partial e^{ip\cdot X}$ and anti-holomorphic counterparts become zero
based on the calculation from (\ref{eq:Chdef}) and (\ref{eq:ehcal}).
Therefore, the fourth term is obtained by the correlator of $e(h)+\frac{1}{8\pi}e^{ip\cdot X}$
in (\ref{eq:delBPsi20}).
Since ${\cal F}$ is defined by (\ref{eq:calF})
and $f(z)$ in (\ref{eq:delBPsi20}) is given by (\ref{eq:zmap}),
we find that
\begin{eqnarray}
 \partial {\cal F}(z_0,\bar{z}_0)=
\bar{\partial} {\cal F}(z_0,\bar{z}_0)=\frac{i}{\mathrm{Im} \,z_0}.
\end{eqnarray}
Combining the results, 
the fourth term of (\ref{eq:Adilaton}) is given by
\begin{eqnarray}
 i(2\pi)^{26}\delta^{26}(p)\,
\frac{1}{2\sqrt{24}}\frac{\sqrt{\pi}}{16}(4\pi^2\alpha')^{-7}
\,\left(4\sum_{k=0}^\infty \frac{25+k}{25\cdot 24}+2\right).
\label{eq:amp3}
\end{eqnarray}
Here, the same divergent series as seen in (\ref{eq:amp2}) arises due to
the correlation involving $e(h)$.  However, both series in
(\ref{eq:amp2}) and (\ref{eq:amp3}) mutually cancel each out.  This
cancellation can always be achieved by applying the same regularization
scheme to the correlator of $e(h)$ in both terms. Furthermore, it is
worth noting that the cancellation depends on the specific form of the
conformal mapping from the unit disk to the upper-half plane, but the
outcome remains independent of the parameters, $\theta_0$, $\theta_1$
and $r_0$ in the mapping.

Finally, the dilaton tadpole can be obtained by evaluating (\ref{eq:Adilaton})
in the BRST formalism.  By incorporating
 (\ref{eq:amp1}), (\ref{eq:amp2}) and
(\ref{eq:amp3}), the result is given by\footnote{
We note that this is also obtained from the effective action for the dilaton field $\tilde{\Phi}$ in the Einstein metric from (8.7.24) with $p=25$, (8.7.23a) with $D=26$ and (8.7.26) in \cite{Polchinski:1998rq} as
\begin{eqnarray}
-i\tau_{25}\frac{14}{12}\frac{\sqrt{24}\kappa}{2}(2\pi)^{26}\delta^{26}(p)=
 -i(2\pi)^{26}\delta^{26}(p)\,
\frac{26+2}{2\sqrt{24}}\,\frac{\sqrt{\pi}}{16}(4\pi^2\alpha')^{-7}
\end{eqnarray}
up to a sign ambiguity.
}
\begin{eqnarray}
 {\cal A}=
 i(2\pi)^{26}\delta^{26}(p)\,
\frac{26+2}{2\sqrt{24}}\,\frac{\sqrt{\pi}}{16}(4\pi^2\alpha')^{-7}.
\end{eqnarray}
Here, $26+2$ in the numerator is a well-known result. Initially, it was
determined as $24$ in \cite{Douglas:1986eu}, but subsequent discussions
in \cite{Liu:1987nz} led to the conclusion that it should be replaced
with $26+2$ to account for the contribution of the Euler characteristic
of the world-sheet.  It's essential to note that even though the Euler
characteristic is calculated through the integration of curvature, the
value $26+2$ is obtained from local operators within the BRST formalism.

\section{Concluding remarks
\label{sec:remarks}}

In this paper, using the FP procedure for the gauge fixing of the CKG of  a disk,
we have constructed closed string vertex operators with various ghost numbers
from matter primary fields $V(z,\bar{z})$ with the conformal weight $(1,1)$ and ghosts $c(z),\tilde{c}(\bar{z})$.
These operators can be treated as $n$-forms $\omega_n^{~g}$ ($n=0,1,2$)
with respect to  $dz$ and $d\bar{z}$,
where $g$ denotes the ghost number.
We have found that these $\omega_n^{~g}$ give solutions to the descent equations:
\begin{eqnarray}
&{\bm \delta}_{\rm B}\omega_2^{~g}=-d\omega_1^{~g+1},
\qquad
{\bm \delta}_{\rm B}\omega_1^{~g}=d\omega_0^{~g+1},
\qquad
{\bm \delta}_{\rm B}\omega_0^{~g}=0,
\end{eqnarray}
in the BRST formalism.
In particular, $\omega_n^{~g}$ with $n+g=3$ can be obtained using
the factor $\partial c-\bar{\partial} \tilde{c}$ 
from $\omega_n^{~g}$ with $n+g=2$.
We note that the constructed $\omega_n^{~g}$ are all conformal invariant up to ${\bm \delta}_{\rm B}$-exact and $d$-exact terms.

Similarly, we have constructed solutions to the descent equations associated with the dilaton vertex operator, which are denoted as $\Omega_n^{~g}$ ($n+g=2,3$).
They are based on the matter field $\partial X\cdot \bar{\partial} X e^{ip\cdot X}$,
which is not primary for nonzero momentum, and are covariant with respect to the 26-dimensional space-time.
We have explicitly shown that $\Omega_n^{~g}$ is invariant up to  ${\bm \delta}_{\rm B}$-exact and $d$-exact terms under the conformal transformations.
It is characteristic that $\Omega_0^{~2}$ includes 
the ghost dilaton $c\partial^2c -\tilde{c}\bar{\partial}^2\tilde c$
and $\Omega_0^{~3}$ is made of an infinite sum involving $X\cdot \partial X e^{ip\cdot X}$, $X\cdot \bar{\partial} X e^{ip\cdot X}$.
Using $\Omega_0^{~3}$ and its conformal transformation, we have evaluated the dilaton tadpole amplitude on the disk and obtained the correct result with the factor $26+2$, which was calculated including a contribution from the curvature of the world-sheet in the previous literature.
In our computation, we have found that there is a non-zero contribution from the ${\bm \delta}_{\rm B}$-exact term due to $X^{\mu}$.

As a closed string version of the calculations in \cite{Seki:2019ycz, Seki:2021ivm, Kishimoto:2021csf},
we can use the above $\omega_n^{~g}$ to evaluate the closed string amplitudes with the mostly BRST exact operator.
We will describe the details of this issue elsewhere.

\section*{Acknowledgments}
We would like to thank Tomomi Kitade for useful discussions in the early
stage of the work.  This work was supported in part by JSPS Grant-in-Aid
for Scientific Research (C) \#20K03972.  I.~K.~was supported in part by
JSPS Grant-in-Aid for Scientific Research (C) \#20K03933.  S.~S.~was
supported in part by MEXT Joint Usage/Research Center on Mathematics and
Theoretical Physics at OCAMI and by JSPS Grant-in-Aid for Scientific
Research (C) \#17K05421 and \#22K03625. T. T. was supported in part by
JSPS Grant-in-Aid for Scientific Research (C) \#23K03388.

\bibliographystyle{utphys}
\bibliography{referencev4}

\end{document}